\documentstyle[aps,multicol,epsf,epsfig,float]{revtex}

\def\CC{{\rm\kern.24em \vrule width.04em height1.46ex depth-.07ex
\kern-.30em C}}
\def\P{{\rm I\kern-.25em P}}
\def\RR{{\rm
         \vrule width.04em height1.58ex depth-.0ex
         \kern-.04em R}}

\def\RR{{\rm\kern.24em \vrule width.04em height1.46ex depth-.07ex
\kern-.30em R}}
\def\P{{\rm I\kern-.25em P}}
\def\RR{{\rm
         \vrule width.04em height1.58ex depth-.0ex
         \kern-.04em R}}

\newcommand{\be}{\begin{equation}}
\newcommand{\ee}{\end{equation}}
\newcommand{\bq}{\begin{eqnarray}}
\newcommand{\eq}{\end{eqnarray}}

\begin{document}
\draft
\title{Decoherence-free dynamical and geometrical entangling phase gates}
\author{Jiannis K. Pachos\footnote{jiannis.pachos@imperial.ac.uk} and Almut Beige}
\address{Blackett Laboratory, Imperial College London, Prince Consort Road, London,
SW7 2BW, UK}
\date{\today}
\maketitle

\begin{abstract}
It is shown that entangling two-qubit phase gates for quantum computation with atoms
inside a resonant optical cavity can be generated via common laser addressing,
essentially, within one step. The obtained dynamical or geometrical phases are
produced by an evolution that is robust against dissipation in form of spontaneous
emission from the atoms {\em and} the cavity and demonstrates resilience against
fluctuations of control parameters. This is achieved by using the setup introduced
by Pachos and Walther [Phys. Rev. Lett. {\bf 89}, 187903 (2002)] and employing
entangling Raman- or STIRAP-like transitions that restrict the time evolution of the
system onto stable ground states.
\end{abstract}
\vspace*{0.2cm}
\noindent

\begin{multicols}{2}

\section{Introduction}

The phase of a state vector is undoubtedly one of the central curiosities that
differentiate quantum from classical mechanics. Especially, the presence of quantum
phases in non-intuitive interference experiments has been the focus of intellectual
excitement in many studies (see for example
\cite{Scully,hradil,bouw,brezger,Berry}). The presence of a phase is even more
striking in a state that exhibits entanglement. In particular, the research on
highly correlated subsystems has contributed to the boosting of innovative
technological applications \cite{Dowling,Dowling2,Bennett}. Such an example is
quantum computation where the entangling phase gate often provides the basic
two-qubit gate -- an essential ingredient for the realization of arbitrary quantum
algorithms. This gate corresponds to a unitary operation that changes the phase of
one subsystem depending on the state of another one. In the last years, considerable
effort has been made to find efficient ways for its experimental implementation.

For a proposed gate implementation to be feasible with present technology, it is
important that the scheme is widely independent from various control parameters like
the operation time. Minimizing the control errors will augment the efforts for
engineering scalable quantum information processing with high accuracy.  Another
problem arises from the fact that it is very difficult to isolate a quantum
mechanical system completely from its environment without loosing the possibility to
manipulate its state. Uncontrollable environmental couplings lead in general to
dissipation and the loss of information. In addition, the phase of a state vector is
very fragile with respect to decoherence. As a result, phase factors are rather hard
to generate and to store in engineered systems.

This paper analyzes in detail entangling phase operations that are especially
designed to bypass the dissipation problem and guarantee very high fidelities for a
wide range of experimental parameters. In particular, we consider the quantum
behaviour of atoms inside an optical cavity which has been observed experimentally
by Hennrich {\em et al.} \cite{Rempe1} and, more recently, by J. McKeever {\em et
al.} \cite{McKeever} and by Sauer {\em et al.} \cite{Sauer}. The scheme presented
here involves two atoms trapped at fixed positions inside an optical cavity and can
be implemented using the technology of the recent calcium ion experiment
\cite{wolfgang}. Alternatively, neutral atoms can be trapped with a standing laser
field \cite{Schoen}, as in the experiment by Fischer {\em et al.}
\cite{Rempe2}, or in an optical lattice \cite{Sauer,horak}. In the last decade,
several atom-cavity quantum computing schemes have been proposed
\cite{Pellizzari,domokos,beige,zheng,jianni,jane,Recati,you,sorensen},
each of them having its respective merits.

To implement quantum phase gates we utilize adiabatic processes that result in the
generation of dynamical or geometrical phases \cite{Berry2}. Employing geometrical
phases is an intriguing way of manipulating quantum mechanical systems. They exhibit
independence from the operation time of the control evolution and are robust against
perturbations of system parameters as long as certain requirements, concerning the
geometrical characteristics of the evolution, are satisfied. To avoid dissipation,
we exploit the existence of decoherence-free states
\cite{dfs,dfs2,dfs3,beige2}. As in \cite{jianni}, only {\em ground states}
with no photon in the cavity mode and the atoms in a stable state become populated.
Population transfers between those ground states are achieved with the help of Raman
or STIRAP-like processes \cite{Oreg,Bergmann}. Their control procedures are exactly
the same as for the usual Raman and STIRAP transitions, but now the coupling of the
atoms to the same cavity mode allows for the creation of entanglement. Another
advantage of the proposed scheme is that it can be realized via common laser
addressing and, essentially, within one step.

In the following, each qubit is obtained from two ground states of the same atom.
High fidelities of the final state are achieved even for moderate values of the
atom-cavity coupling constant $g$ compared to the spontaneous cavity decay rate
$\kappa$ and the atom decay rate $\Gamma$. Different from other atom-cavity quantum
computing schemes \cite{Pellizzari,domokos,beige,zheng,jane}, we assume the constant
$g^2$ to be about $100\, \kappa \Gamma$ which is close to the state of the art
technology \cite{Sauer} and should be within the range of experiments in the nearer
future \cite{horak}.

The paper is organized as follows. Section \ref{effective} introduces the
decoherence-free subspace of the system with respect to leakage of photons through
the cavity mirrors. An effective Hamiltonian describing the possible time evolutions
of the system within that subspace is derived. In order to avoid spontaneous
emission from the atoms, we present in Section \ref{passage} different parameter
regimes that minimize the population in the excited atomic levels and result in
entangling Raman and entangling STIRAP transitions. In Section \ref{berry} we employ
the corresponding time evolutions to generate dynamical and geometrical phases for
the realization of two-qubit phase gates for quantum computation. Finally, we
summarize our results in Section \ref{conclusions}.

\section{Elimination of cavity decay} \label{effective}

In the following we consider two atoms (or ions), each of them comprising a
four-level system. A possible level configuration, which is suited for the
implementation of quantum computing with calcium ions, is shown in Figure
\ref{calcium}. Each qubit is obtained from the stable ground states $|0\rangle$ and
$|1\rangle$ of the same atom. In addition, each atom possesses an excited state
$|2\rangle$, that becomes only virtually populated during gate operations, and a
third auxiliary ground state $|\sigma \rangle$. In order to realize an entangling
operation, the two atoms involved should be positioned in the cavity, where both see
the same coupling constant $g$, and laser fields should be applied as shown. As we
see later, the detuning $\delta$ plays an important role in regulating the phase
factor of the final state of the system.

\noindent
\begin{minipage}{3.38truein}
\begin{center}
\begin{figure}[ht]
\centerline{
\epsfig{file=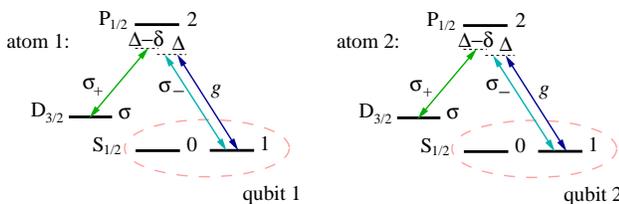,width=3.38truein}
}
\caption[contour]{\label{calcium} Level configuration for atom-cavity quantum computing
with calcium ions showing the driving laser fields. The 2-$\sigma $
and the 2-1 transitions are activated by the laser radiations with
amplitudes and polarizations given by $\Omega_\sigma$, $\sigma_+$ and $\Omega_1$,
$\sigma_-$ respectively.

Each qubit is obtained from two
degenerate $S_{1/2}$ ground states of the same ion.}
\end{figure}
\end{center}
\end{minipage}

Let us denote the cavity photon annihilation and creation operator by $b$ and
$b^\dagger$ while $\Delta$ is the detuning of the cavity with respect to the 2-1
transition of each atom. The occupation of the cavity is then given by $n=\langle
b^\dagger b \rangle$. One laser drives the 2-1 transition of atom $i$ with Rabi
frequency $\Omega^{(i)}_1$ and detuning $\Delta$; another one excites the 2-$\sigma$
transition with Rabi frequency $\Omega^{(i)}_\sigma$ and detuning $\Delta-\delta$.
Going over to the interaction picture with respect to $H_0+\hbar \sum_i
\, [ \, \delta  |\sigma\rangle_{ii} \langle \sigma| + \Delta \, |2
\rangle_{ii} \langle 2| \, ]$ where $H_0$ is the interaction-free Hamiltonian, one
finds
\begin{eqnarray}\label{ham}
H_{\rm I} &=& \sum_{i=1}^2
\hbar g \, \big[ \,|2\rangle_{ii} \langle 1| b + \text{H.c.} \, \big]\nonumber \\
\nonumber \\
&& +{\textstyle { 1 \over 2}} \hbar  \, \big[ \,
\Omega^{(i)}_1 |1\rangle_{ii} \langle 2| + \Omega^{(i)}_\sigma |\sigma\rangle_{ii}
\langle 2| + \text{H.c.} \, \big] \nonumber \\ \nonumber \\
&& - \hbar \delta \, |\sigma\rangle_{ii} \langle \sigma| -
\hbar \Delta \, |2 \rangle_{ii} \langle 2| ~.
\end{eqnarray}
The generation of a non-trivial time evolution requires in general a non-homogeneity
in the Rabi frequencies of the laser fields with respect to the two atoms. Without
loss of generality, we assume here that the Rabi frequencies of laser fields
coupling to the same transition differ only in phase. In the following, entangling
operations are realized by choosing
\begin{eqnarray} \label{rabi}
&& \Omega^{(1)}_\sigma = -\Omega^{(2)}_\sigma \equiv - \Omega_\sigma ~, \nonumber \\
&& \Omega^{(1)}_1 = - \Omega^{(2)}_1 \equiv - \Omega_1 /\sqrt{2}~.
\end{eqnarray}
These parameters can be implemented via common laser addressing by driving each
transition with the same laser field and from an angle that produces a $\pi$ phase
difference between atom 1 and 2.

One of the main sources for decoherence in atom-cavity setups is the leakage of
photons through the cavity mirrors. However, the presence of high spontaneous decay
rates does not necessarily lead to dissipation. As it has been shown in the past,
the presence of a relatively strong atom-cavity coupling constant $g$
\cite{carsten} can have the same effect as continuous measurements whether the
cavity mode is empty or not. As a consequence, the time evolution of the system
becomes restricted onto the {\em decoherence-free} states of the system with respect
to cavity decay. In this Section, we exploit this effect and its consequences for
the time evolution to significantly simplify the requirements for atom-cavity
quantum computation.

In the next Section, Raman and STIRAP-like processes \cite{Oreg,Bergmann} are
introduced in order to minimize the population in the excited atomic state
$|2\rangle$. Assuming that the population in level $2$ is negligible, the only
condition that has to be fulfilled to avoid cavity decay is that the Rabi
frequencies $\Omega_1$ and $\Omega_\sigma$ are sufficiently weak compared to the
atom-cavity coupling constant $g$,
\begin{eqnarray}
\fbox{
$ \label{cond} |\Omega_1| \, ,\,\, |\Omega_\sigma| \ll g $}
\end{eqnarray}
This relation induces two different time scales in the system and allows for the
calculation of the time evolution with the help of an adiabatic elimination. Figure
\ref{level} shows the most relevant transitions if the system is initially prepared
in a qubit state with an empty cavity ($n=0$) and uses the abbreviations
\begin{eqnarray}
&& |\alpha\rangle \equiv \big(\,|12\rangle- |21\rangle\,\big)/\sqrt{2} ~, \nonumber
\\ && |A \rangle \equiv \big( \, |\sigma 1 \rangle +|1 \sigma \rangle \,
\big)/\sqrt{2} ~, \nonumber \\ && |\tilde \alpha \rangle \equiv \big(\, |\sigma 2
\rangle - |2 \sigma \rangle \, \big) /\sqrt{2} ~, \nonumber \\ && |\tilde A \rangle
\equiv \big( \, |\sigma 1 \rangle - |1 \sigma \rangle \, \big)/\sqrt{2} ~.
\end{eqnarray}
Setting the amplitudes of all fast varying states equal to zero and using the
Hamiltonian (\ref{ham}), one finds that the system evolves effectively according to
the Hamiltonian
\begin{eqnarray} \label{eff}
H_{\text{eff}} &=& \big[ \, {\textstyle { 1 \over 2}} \hbar \,
\big( \, \Omega_1 \, |\alpha \rangle \langle 11| + \Omega_\sigma \,
|\alpha \rangle \langle A| + \text{H.c.} \, \big) \nonumber \\ \nonumber \\ && -
\hbar \delta \, |A \rangle \langle A| - \hbar \Delta \, |\alpha\rangle \langle
\alpha| \, \big]
\otimes |0_{\rm cav} \rangle \langle 0_{\rm cav}| ~.
\end{eqnarray}
As expected for adiabatic processes, this Hamiltonian restricts the
time evolution onto a subspace of slowly varying states, which is why
the states $|\tilde{\alpha}\rangle$ and $|\tilde{A}\rangle$ are not
present in Equation (\ref{eff}).
The subspace of the slowly varying states includes, in addition to the ground
states, only one state with population in the excited state $|2 \rangle$. This
state, $|\alpha \rangle |0_{\rm cav} \rangle$, is a zero eigenstate of the
atom-cavity interaction.

\noindent \begin{minipage}{3.38truein}
\begin{center}
\begin{figure}[ht]
\centerline{
\epsfig{file=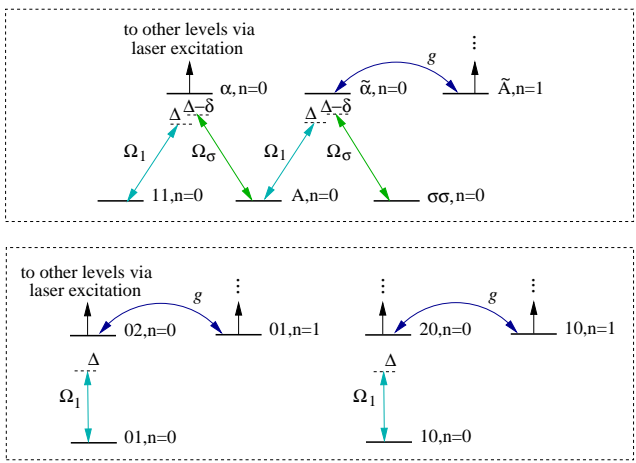,width=3.3truein}}
\vspace{0.2cm}
\caption[contour]{\label{level} Level configuration showing the most relevant
transitions for the initial qubit states $|11\rangle$, $|01\rangle$ and $|10\rangle$
in the presence of mechanisms (see Section \ref{passage}) that minimize the
population in level 2. The presence of a strong cavity coupling
constant $g$ prohibits a transfer of population between
$|A\rangle$ and $|\sigma \sigma\rangle$. This does not effect the
transition between $|11\rangle$ and $|A\rangle$. The atomic state $|00\rangle$
does not see any laser fields and does not change its amplitude in time.}
\end{figure}
\end{center}
\end{minipage}

Note that the Hamiltonian (\ref{eff}) indeed restricts the time evolution of the
system onto its decoherence-free subspace with respect to cavity decay. A state
belongs to this subspace if the resonator mode is in its vacuum state $| 0_{\rm cav}
\rangle$ and the atom-cavity interaction cannot transfer excitation from the atoms
into the resonator. Hence, the decoherence-free subspace includes all ground states
and the state $|\alpha \rangle |0_{\rm cav} \rangle$. They are exactly the slowly
varying states of the system. Furthermore, the effective Hamiltonian can be written
as
\begin{equation}
H_{\rm eff} = I\!\!P_{\rm DFS} \, H_{\rm laser} \, I\!\!P_{\rm DFS}
~,
\end{equation}
where $I\!\!P_{\rm DFS}$ is the projector onto the decoherence-free
subspace
\begin{equation}
I\!\!P_{\rm DFS}=\Big[|\alpha\rangle \langle
  \alpha|+\sum_{i,j=0,1,\sigma} |ij \rangle \langle ij|
\Big] \otimes |0_{\rm cav}\rangle
\langle 0_{\rm cav}|~,
\end{equation}
and $H_{\rm laser}$ is the laser term in (\ref{ham}). One way to
interpret this is to state that condition (\ref{cond}) effectively
induces continuous measurements whether the cavity field is empty or
not \cite{cold}. As a consequence, entanglement can be generated between qubits
via excitation of the maximally entangled state $|\alpha\rangle$.

In the event that condition (\ref{cond}) is not fulfilled, as in all realistic
experiments, the above argumentation does not hold and corrections to the effective
time evolution (\ref{eff}) have to be taken into account. To identify the evolution
of the system more accurately, we use the quantum jump approach
\cite{hegerfeldt,he2,he3} which predicts the no-photon time evolution of the system
with the help of the conditional, non-Hermitian Hamiltonian $H_{\rm cond}$. Given
the initial state $|\psi \rangle$, the state of the system equals $U_{\rm cond}
(T,0) \, |\psi \rangle/\| \cdot \|$ at time $T$ under the condition of no emission.
For convenience, $H_{\rm cond}$ has been defined such that
\begin{equation} \label{P0}
P_0(T) = \| \, U_{\rm cond} (T,0) \, |\psi \rangle \, \|^2
\end{equation}
is the probability for no photon in $(0,T)$. If photons are emitted, then the
computation failed and has to be repeated. To some extent, this can be compensated
by monitoring photon emissions with good detectors. Alternatively, quantum
teleportation can be employed to perform a whole algorithm by selecting the
successful gates \cite{Knill}.

Let us now consider a concrete example to see how well the elimination of cavity
decay works. For the two four-level atoms in Figure \ref{calcium} it is
\cite{jianni}
\begin{eqnarray}\label{xxx}
H_{\rm cond} &=& H_{\rm I} - {\textstyle{{\rm i} \over 2}} \hbar \kappa \, b^\dagger
b - {\textstyle{{\rm i} \over 2}} \hbar \Gamma \, \sum_{i=1}^2  |2 \rangle_{ii}
\langle 2| ~.
\end{eqnarray}
Suppose that the system is initially prepared in $|01 \rangle |0_{\rm cav}
\rangle$ and a laser field is applied with $\Omega_1 \neq 0$ that excites the 2-1
transition of both atoms for a significant time $T$. Equation (\ref{eff}) then
implies the inhibition of any time evolution since $|01\rangle|0_{\rm cav} \rangle$
is a zero eigenstate of $H_{\rm eff}$.  Indeed, the numerical solution of the time
evolution given by the Hamiltonian (\ref{xxx}) reveals that the final state
coincides with the initial state of the system with fidelity $F \equiv 1$ under the
condition of no photon emission in $(0,T)$.  The success rate $P_0(T)$  as a
function of $\Delta$ and $\Omega_1$ is shown in Figure \ref{ramanP}. As expected,
$P_0$ is the closer to one the smaller the Rabi frequency $\Omega_1$ and its size is
widely independent from the detuning $\Delta$.

\noindent \begin{minipage}{3.38truein}
\begin{center}
\begin{figure}[ht]
\centerline{
\epsfig{file=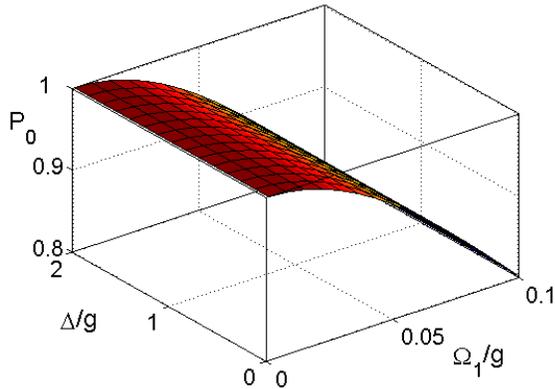,width=3.00truein}}
\caption[contour]{\label{ramanP} The success rate $P_0(T)$ as a function of the
Rabi frequency $\Omega_1$ and the detuning $\Delta$ for the initial state
$|01\rangle |0_{\rm cav} \rangle$, $\kappa =\Gamma=0.1\, g$, $\delta =0$ and
$T=2000/g$. If one allows for a short transition time at the end of the operation
for unwanted states to decay, then the final state of the system equals exactly the
initial one.}
\end{figure}
\end{center}
\end{minipage}

\section{Elimination of atom decay and entangling operations} \label{passage}

For optical cavities, the atom decay rate $\Gamma$ is in general of about the same
size as the parameters $g$ and $\kappa$. We therefore still need to overcome the
problem of spontaneous emission from the atoms in order to realize a coherent time
evolution. This can be achieved by minimizing the population in level $2$ and
keeping the system effectively in the ground states $|11\rangle |0_{\rm cav}
\rangle$ and $|A\rangle  |0_{\rm cav} \rangle$. To realize this we employ in the
following Raman or STIRAP-like transitions \cite{Oreg,Bergmann}. These are directly
applicable since the relevant states comprise a $\Lambda$-type three-level
configuration. Entanglement between the atoms is created by coupling $|11 \rangle$
to the maximally entangled state $|A\rangle$.

\subsection{Entangling Raman transitions} \label{raman}

One way to implement transitions without populating the excited state $|\alpha
\rangle$ is to choose the detuning $\Delta$ much larger than the Rabi frequencies
$\Omega_1$ and $\Omega_\sigma$,
\begin{equation} \label{cond3}
\fbox{
$|\Omega_1| \, ,\,\, |\Omega_\sigma| \ll  \Delta ~$}
\end{equation}
This choice of parameters results in the realization of a Raman-like transition
since the large detuning $\Delta$ introduces an additional time scale in the time
evolution (\ref{eff}). Adiabatically eliminating the excited state $|\alpha \rangle$
reveals that the system is effectively governed by the Hamiltonian
\begin{eqnarray} \label{red}
H_{\rm eff} &=& {\textstyle {1 \over 2}} \hbar \Omega \, \big[ \, |11 \rangle
\langle A| + \text{H.c.}  \, \big]  \nonumber \\ \nonumber \\ && - \hbar \Delta_{11}
\, |11 \rangle \langle 11| -\hbar \Delta_A \, |A \rangle \langle A|
\end{eqnarray}
with the abbreviations
\begin{eqnarray}
\Omega \equiv {\Omega_1 \Omega_\sigma^* \over 2 \Delta}~,~
\Delta_{11} \equiv - {|\Omega_1|^2 \over 4 \Delta} ~,~
\Delta_A \equiv \delta - {|\Omega_\sigma|^2 \over 4 \Delta}  ~.
\end{eqnarray}
Solving the time evolution for the case where the lasers are turned on
for a period $T$ with constant amplitude yields
\end{multicols}
\begin{eqnarray} \label{redev}
U_{\rm eff}(T,0) &=& \exp \left( {\textstyle {{\rm i} \over 2}} (\Delta_{11}
+\Delta_A) T \right) \left[ \, \left( \, \cos {\textstyle {K T \over 2}} +
{\textstyle{{\rm i}(\Delta_{11} -\Delta_A) \over K}} \, \sin {\textstyle {K T \over
2}} \right) |11\rangle \langle 11| \right. \nonumber \\ && \left.
-{\textstyle{\Omega
\over K} }\sin {\textstyle{K T \over 2}} \, \big( \, |11 \rangle \langle A| + |A\rangle
\langle 11| \, \big) +\left( \, \cos {\textstyle {K T \over 2}} -{\textstyle {{\rm
i}(\Delta_{11} -\Delta_A) \over K}} \, \sin {\textstyle {K T \over 2}} \right)
|A\rangle \langle A| \, \right] ~,
\end{eqnarray}
\begin{multicols}{2}
\noindent
where
\begin{equation}
K\equiv \left(|\Omega|^2 +(\Delta_{11} -\Delta_A)^2
\right)^{1/2}
\end{equation}
plays the role of a Rabi frequency. As the evolution (\ref{redev}) can be used to
create entanglement, we call it an entangling Raman (E-Raman) transition.

In order to see how well the elimination of the dissipative states works, let us
consider as an example the preparation of the maximally entangled state $|A
\rangle$. If the atoms are initially in $|11\rangle$, this can be achieved by
applying a laser pulses of length $T= \pi /K = 2\pi \Delta |\Omega_1|^2$ with
$\delta=0$ and $\Omega_\sigma = \Omega_1$. Figures \ref{ramanF} and \ref{ramanP0}
present the fidelity $F$ and success rate $P_0$, respectively, of this entangling
operation as a function of the Rabi frequency $\Omega_1$ and the detuning $\Delta$.
They result from a numerical integration of the time evolution with the full
Hamiltonian (\ref{xxx}).

In particular for $\kappa=\Gamma=0.1 \, g$, the fidelity of the final state can be
as high as 0.993 if one chooses $\Omega_1=0.01 \, g$ and $\Delta=1.357 \, g$ and
allows the success probability to be as high as $P_0=0.857$. Higher fidelities can
only be obtained in the presence of smaller decay rates. In general one can improve
the success rate $P_0$ by a few percent by increasing the detuning $\Delta$ and by
sacrificing an error from the maximal fidelity of the order of $10^{-3}$. While the
fidelity exhibits a maximum as a function of $\Delta$, the success probability
increases monotonically with $\Delta$. For large detunings, the adiabatic
elimination of the cavity mode is no longer efficient (see Figure \ref{level})
resulting in the reduction of $F$. Varying the Rabi frequency gives larger
fidelities for smaller $\Omega_1$ since the adiabatic elimination of cavity
excitation as well as the elimination of $| \alpha \rangle$ becomes more efficient.
However, the smaller $\Omega_1$ the larger the operation time $T$ of the entangling
evolution.

\noindent \begin{minipage}{3.38truein}
\begin{center}
\begin{figure}[ht]
\centerline{
\epsfig{file=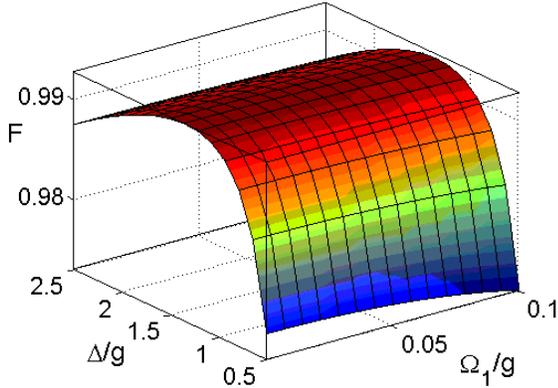,width=3.00truein}
}
\caption[contour]{\label{ramanF} The fidelity $F$ for the preparation of the
maximally entangled state $|A\rangle$ given the initial state $|11\rangle$ as
functions of the Rabi frequency $\Omega_1$ and the detuning $\Delta$ for
$\Omega_\sigma=\Omega_1$, $\Gamma=\kappa=0.1\,g$ and $\delta=0$.}
\end{figure}
\end{center}
\end{minipage}

\noindent \begin{minipage}{3.38truein}
\begin{center}
\begin{figure}[ht]
\centerline{
 \epsfig{file=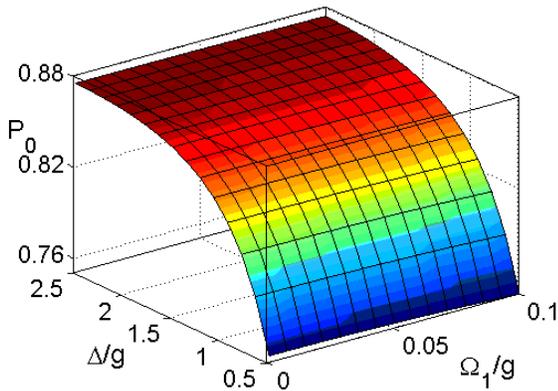,width=3.00truein}
}
\caption[contour]{\label{ramanP0} The success rate $P_0$ for the preparation of
the maximally entangled state $|A\rangle$ for the same parameters as in Figure
\ref{ramanF}.}
\end{figure}
\end{center}
\end{minipage}

\subsection{Entangling STIRAP processes} \label{sirup2}

Another way to transfer population between the ground states $|11 \rangle$ and $|A
\rangle$ without ever populating the excited state $|\alpha \rangle$ is to employ a
STIRAP-like process \cite{Oreg,Bergmann}. As this is an entangling process we can
call it an entangling STIRAP (E-STIRAP) transition. To see how the suppression of
spontaneous emission from the atoms works in this case let us consider again the
effective Hamiltonian (\ref{eff}) and note that its eigenvalues for $\delta \ll
\Delta$ are
\begin{equation}
E_0 = 0 ~, ~ E_{\pm} = {\textstyle {1 \over 2}} (-\Delta \pm
\sqrt{|\Omega_\sigma|^2+ |\Omega_1|^2+\Delta^2}) ~.
\end{equation}
For convenience, we now introduce the interaction picture with respect to the
Hamiltonian $H_0=-\hbar \delta |A\rangle \langle A|$. Within that picture, the
eigenvectors of the effective Hamiltonian become time dependent and the dark state
of the system equals
\begin{equation}
|E_0 \rangle \equiv {1 \over \sqrt{|\Omega_\sigma|^2+ |\Omega_1|^2}}\,
\big(\Omega_\sigma |11\rangle -{\rm e}^{-{\rm i} \delta t}\,\Omega_1|A\rangle\big)
~.
\end{equation}
The other two eigenstates $|E_{\pm} \rangle$ occupy the antisymmetric state $|\alpha
\rangle$ and hence possess population in the excited atomic state $|2\rangle$.

Spontaneous emission from the atoms can therefore be avoided if the system remains
constantly in the dark state $|E_0 \rangle$. Nevertheless, a nontrivial entangling
time evolution between the ground states $|11\rangle$ and $|A\rangle$ can be
induced, by varying the laser amplitudes $\Omega_1$ and $\Omega_\sigma$ in time. If
this happens slow compared to the time scale given by the eigenvalues $E_\pm$, no
population is transferred into the eigenstates $|E_\pm
\rangle$. Assuming adiabaticity, the system follows the changing parameters and
remains in the dark state. The fact that $|E_0
\rangle$ incorporates a time dependent phase factor can be used to induce a certain
phase on the final state $|A\rangle$ by choosing the small laser detuning $\delta$
appropriately.

In the following we assume $\Delta =0$ since this maximizes the energy distance
between $|E_0 \rangle$ and its nearest eigenstate. For convenience, we also consider
the case where $\Omega_1$ and $\Omega_\sigma$ are real and define
$
\tan\theta \equiv \Omega_1/\Omega_\sigma$ and $\phi \equiv -\delta t$. The
eigenstate corresponding to the zero eigenvalue then takes the familiar form
\begin{equation}
|E_0  \rangle =\cos \theta \, |11\rangle - {\rm e}^{{\rm i} \phi} \sin \theta \,
|A\rangle ~.
\end{equation}
The adiabatic condition is satisfied if
\begin{equation}
\fbox{
$|\dot{\theta}| \, , \,\, |\dot{\phi}| \ll \sqrt{\Omega_\sigma ^2 + \Omega_1^2}$}
\label{adia}
\end{equation}
If the time dependence of $\theta$ and $\phi$ is indeed such that the initial state
of the system coincides with the dark state $|E_0(0) \rangle$, the effective
evolution along a path $C$ in the $(\theta,\phi)$ parameter space can easily be
calculated. As a consequence of the adiabatic theorem, the effective evolution is
diagonal with respect to the basis given by the eigenvectors $|E_0 \rangle$ and
$|E_\pm \rangle$. Especially for the case where $\theta(0) = 0$ and with respect to
the basis states $|E_0(0) \rangle$,  $|E_+(0) \rangle$ and  $|E_-(0) \rangle$, the
effective time evolution of the system can be written as
\begin{equation} \label{sirup}
U_{\rm eff}(T,0)=R(\theta,\phi) \left(
\begin{array}{ccc} {\rm e}^{{\rm i} \varphi_{0}}  & 0 & 0 \\
0 & {\rm e}^{{\rm i} \varphi_{+}} & 0 \\ 0 & 0 & {\rm e}^{{\rm i} \varphi_{-}}
\end{array} \right)
\end{equation}
with $\theta = \theta(T)$, $\phi = \phi(T)$ and
\begin{eqnarray}
&& R(\theta,\phi) = {\textstyle {1 \over 2}} \times \nonumber \\
&&
\left( \begin{array}{ccc}
2 \cos \theta  & \sqrt{2}\sin \theta & \sqrt{2}\sin \theta \\
- \sqrt{2}{\rm e}^{{\rm i} \phi}  \sin \theta & (1+{\rm e}^{{\rm i}\phi}
\cos
\theta) & -(1 - {\rm e}^{{\rm i}\phi}  \cos \theta) \\
\sqrt{2}{\rm e}^{{\rm i} \phi}  \sin \theta & (1-{\rm e}^{{\rm i}\phi} \cos
\theta) & -(1 + {\rm e}^{{\rm i}\phi}  \cos \theta)
\end{array} \right) ~\!\!.
\end{eqnarray}
The phases $\varphi_i \equiv \varphi^{\text{d}}_i +\varphi^{\text{g}}_i$ ($i=0, \,
\pm$) are in general the sum of a dynamical and a geometrical phase with
\begin{equation} \label{dyn}
\varphi^{\text{d}}_{0}=0 ~,~ \varphi^{\text{d}}_{+}= - \varphi^{\text{d}}_{-}
= - {\textstyle {1 \over 2}} \int_0^T \sqrt{\Omega_1^2 +\Omega_\sigma^2} \, {\rm d}t
\end{equation}
while the geometrical phases equal
\begin{equation} \label{geo}
\varphi^{\text{g}}_{0} = \oint_C \sin^2 \theta \, {\rm d} \phi
~,~ \varphi^{\text{g}}_{+} = \varphi^{\text{g}}_{-} = {\textstyle {1 \over 2}}
\oint_C \cos^2 \theta \, {\rm d}\phi
\end{equation}
for a closed circular loop $C$.

As an example, let us consider the case where the initial state $|11 \rangle$ is
transferred into the maximally entangled state $|A\rangle$. This can be achieved by
varying the Rabi frequencies $\Omega_1$ and $\Omega_\sigma$ slowly in a so-called
counterintuitive pulse sequence. Figure \ref{double} and \ref{double1} show the
result of a numerical solution of the corresponding time evolution taking the full
Hamiltonian (\ref{xxx}) into account. In particular, we calculated the fidelity and
the success of the scheme as a function of the maximum laser amplitude $\Omega$ and
the frequency $\omega$ for $\Delta = \delta=0$, $\kappa =\Gamma =0.1 \, g$ and
\begin{equation}
\Omega_\sigma(t) =
\left\{ \begin{array}{cl} \Omega \, \sin \omega t \, , & {\rm for} ~ t \in \big(0,
{\textstyle {2 \over 3}} T \big) \\[0.2cm]
0 \, , & {\rm for} ~ t \in \big({\textstyle {2 \over 3}} T,T \big) \end{array}
\right.
\end{equation}
and
\begin{equation}
\Omega_1(t) = \left\{ \begin{array}{cl} 0 \, , & {\rm for} ~ t \in \big(0,
{\textstyle {1 \over 3}} T \big) \\[0.2cm]
\Omega \, \sin \omega \big( t - {\textstyle {1 \over 3}T} \big) \, , &
{\rm for} ~ t \in \big( {\textstyle {1 \over 3}} T , T \big)
\end{array} \right.
\end{equation}
with the total operation time $T = 3 \pi /(2 \omega)$.

\noindent \begin{minipage}{3.38truein}
\begin{center}
\begin{figure}[ht]
\centerline{
 \epsfig{file=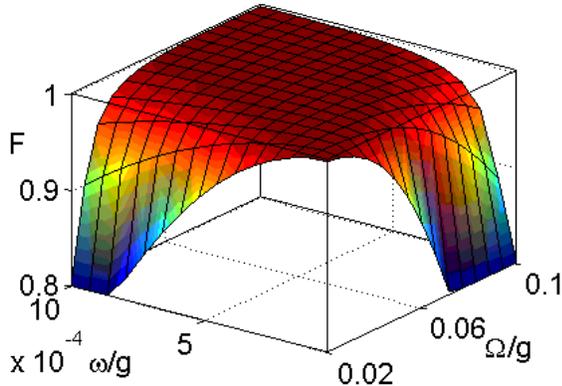,width=3.00truein}
}
\caption[contour]{\label{double} The fidelity $F$ for the preparation of the maximally
entangled state $|A \rangle$ as a function of the maximum Rabi frequency $\Omega$
and frequency $\omega$ for $\Delta = \delta=0$ and $\kappa=\Gamma=0.1\, g$.}
\end{figure}
\end{center}
\end{minipage}

\noindent \begin{minipage}{3.38truein}
\begin{center}
\begin{figure}[ht]
\centerline{
\epsfig{file=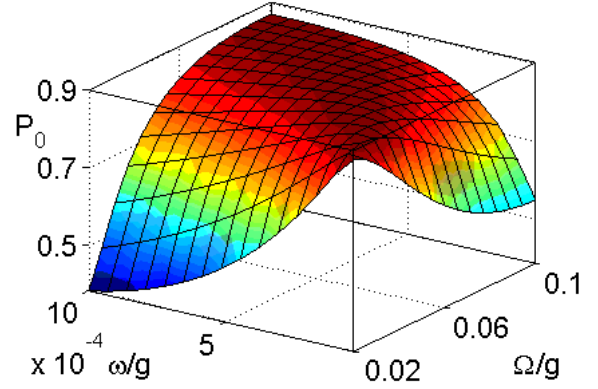,width=3.00truein}
}
\caption[contour]{\label{double1} The success rate $P_0$ for the proposed
entangled state preparation scheme for the same parameters as in Figure
\ref{double}.}
\end{figure}
\end{center}
\end{minipage}

In contrast to E-Raman processes, the fidelity $F$ of the final state can, in
principle, be arbitrarily close to one independent of the size of the spontaneous
decay rates $\kappa$ and $\Gamma$. For example, for $\Omega=0.02 \, g$ and $\omega =
4 \cdot 10^{-5} \, g$ the fidelity becomes $F=0.998$ and corresponds to the success
rate $P_0 = 0.876$  which is still within acceptable limits. In general, a large
$\Omega$ leads to the population of the excited state $|2\rangle$ which might result
in the emission of a photon with rate $\Gamma$ or in the population of the cavity
mode followed by the leakage of a photon through the cavity mirrors. As a result,
$F$ and $P_0$ decrease for increasing laser amplitudes $\Omega$. On the other hand,
for very weak Rabi frequencies $(\Omega \approx 0)$ the fidelity and the success
rate increase as $\omega$ decreases which results in a longer operation time $T$.
This is in agreement with condition (\ref{adia}) which implies that the adiabatic
evolution is more successful for slower evolutions. Nevertheless, the success rate
$P_0$ cannot be increased arbitrarily. The reason is that very long operation times
unavoidably lead to the population of the cavity mode in the presence of a finite
value for $\Omega$ and the elimination of cavity decay does no longer hold.

\section{Decoherence-free dynamical and geometrical phase gates} \label{berry}

In this Section we employ the decoherence-free E-Raman and E-STIRAP processes
introduced in the last Section for the realization of two-qubit quantum gate
operations. In particular, we consider the implementation of the controlled phase
gate
\begin{equation} \label{gate}
CP={\text{diag}} \, \big(1,1,1,{\rm e} ^{{\rm i} \varphi} \big) ~,
\end{equation}
where the qubit state $|11\rangle$ collects the phase $\varphi$ while the other
qubit states remain unaffected. This gate constitutes, together with a general
single-qubit rotation, a universal set of gates and allows for the implementation of
any desired unitary time evolution. Attention is paid to their stability against
fluctuations of most system parameters and we analyze in detail the efficiency of
various processes.

\subsection{Raman-based Dynamical Phase Gates}

Let us begin with the description of a possible implementation of (\ref{gate}) with
the help of the E-Raman process  introduced in Section \ref{passage}A. From Equation
(\ref{redev}) one sees that an applied laser pulse of length $T$ evolves the initial
state $|11\rangle$ according to
\begin{eqnarray}
U_{\text{eff}}(T,0)|11\rangle &=& \exp \left( {\textstyle {{\rm i} \over 2}}
(\Delta_{11} +\Delta_A)  T \right) \times \nonumber \\ \nonumber \\ &&
\hspace*{-2.5cm}  \Big[ \, \Big( \cos {\textstyle{K T \over 2}} + {\rm i} {
\textstyle{\Delta_{11} -\Delta_A \over K}}
\sin {\textstyle{K T \over 2}} \Big) |11 \rangle
- \textstyle{\Omega \over K} \sin {\textstyle{K T \over 2}} |A\rangle \, \Big]   ~ ,
\nonumber \\ \label{11}
\end{eqnarray}
while the amplitudes of the qubit states $|00 \rangle$, $|01 \rangle$ and $|10
\rangle$ do not change in time.  The desired phase gate (\ref{gate}) can therefore
be realized in the minimal time if one chooses
\begin{equation} \label{time}
T = {2 \pi / K} ~.
\end{equation}
For this parameter choice, the sine term in (\ref{11}) vanishes and the cosine term
equals $-1$ giving finally
\begin{equation}
\varphi = \pi + {\textstyle {1 \over 2}} (\Delta_{11} + \Delta_A ) T ~.
\end{equation}
Again, the phase $\varphi$ can be controlled by adjusting the size of the detuning
$\delta$. Especially, for
\begin{equation}
\delta = {\textstyle {1 \over 4 \Delta}} \big( |\Omega_1|^2 + |\Omega_\sigma|^2 \big)
\end{equation}
one obtains $\varphi=\pi$ and the amplitude of the state $|11\rangle$ accumulates a
minus sign. This particular evolution is illustrated in Figure \ref{Raman} using a
Bloch's sphere representation. Note that the condition of timing the evolution such
that the sine term becomes zero, is robust against first order timing errors in $T$
due to the behaviour of the sine function around $\pi$.

\noindent \begin{minipage}{3.38truein}
\begin{center}
\begin{figure}[ht]
\centerline{
\epsfig{file=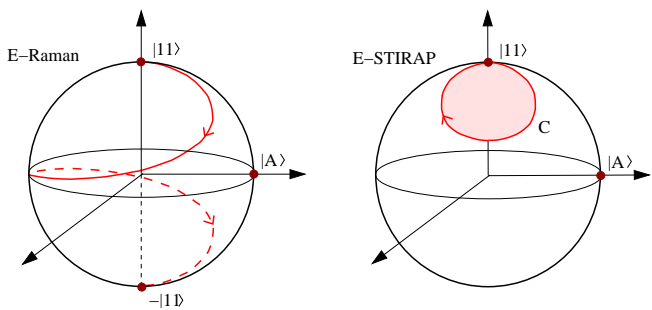,width=3.38truein}
\vspace{0.3cm}
}
\caption[Raman]{\label{Raman} The E-Raman and E-STIRAP evolutions depicted on the Bloch
sphere. In the E-Raman evolution, the generation of the phase factor in front of
$|11\rangle$ is explicitly shown. In the E-STIRAP evolution, the acquired phase
relates to the solid angle spanned by the path $C$ on the sphere.}
\end{figure}
\end{center}
\end{minipage}

To minimize the experimental effort for the realization of the phase gate
(\ref{gate}), one should notice that  a non-trivial E-Raman transition can also be
induced using only one laser field coupling to the 1-2 transition of each atom.
Indeed, for $\Delta_A =0$ and the laser pulse duration $T=\pi/K$, Equation
(\ref{11}) yields
\begin{equation}
U_{\text{eff}}(T,0)|11\rangle = - \exp \big( {\rm i} \Delta_{11} T \big) \,
|11\rangle ~.
\label{onelaser}
\end{equation}
For this case, Figure \ref{level} shows that the initial state $|11\rangle$ couples
only to the excited state $|\alpha \rangle$ via the applied heavily detuned laser
field.  To see that the phase operation (\ref{onelaser}) is robust against
fluctuations of $\Omega_1$, we now consider the Rabi frequency being time dependent.
In this case the phase accumulated by the initial state $|11 \rangle$ becomes
$\varphi =\pi + \int_0 ^T \Delta_{11} {\rm d}t$. This clearly shows that statistical
perturbations of $\Omega_1$ average out due to dependence of the acquired phase on
the integral over $\Delta_{11}$ along $(0,T)$.

\subsection{STIRAP-based Geometrical Phase gate}

Another possibility to implement the controlled two-quit phase gate (\ref{gate}) is
provided by the E-STIRAP process. Let us first consider the case of the generation
of a dynamical phase as in \cite{jianni}. To do so we vary $\theta$ from zero to
$\pi$, then apply a $2 \pi$ pulse to transform the state $|\sigma \rangle$ to
$-|\sigma \rangle$ for both atoms and then reduce $\theta$ back to zero. At the end
of this time evolution, the state $|11\rangle$ acquires an overall minus sign, while
the other qubit states $|00\rangle$, $|01\rangle$ and $|10\rangle$ remain unchanged.
The result is the conditional phase gate with $\varphi=\pi$.

Alternatively, a geometrical phase can be generated by continuously varying the
parameters $\theta$ and $\phi$ in a cyclic fashion \cite{pachos1,Bouwmeester}.
Starting from $\theta=0$ one can perform a cyclic adiabatic evolution on the
$(\theta,\phi)$ plane along a loop $C$.  As predicted by Equations
(\ref{sirup})-(\ref{geo}), the qubit state $|11\rangle$ then acquires
the geometrical phase
\begin{equation}
\varphi = \varphi^{\rm g}_0 = \oint_C \sin^2 \theta \, {\rm d}\phi
\end{equation}
while the dynamical phase $\varphi^{\rm d}_0$ is identically zero. The concrete size
of $\varphi$, which characterizes the phase gate (\ref{gate}), depends therefore
only on the solid angle spanned by the path $C$ on the Bloch sphere (see Figure
\ref{Raman}). Hence it is very robust against statistical fluctuations of the Rabi
frequencies of the applied laser fields and it is independent of the total gate
operation time.

Let us now consider a concrete laser configuration, where the laser field
$\Omega_\sigma$ is turned on and kept constant. To perform a non-trivial loop in the
control parameter space $(\theta,\phi)$ starting from the $(0,0)$ point we proceed
as follows. As described in Section \ref{sirup2}, a non-zero detuning $\delta$
results in the accumulation of a phase $\phi=-\delta t$ when we introduce a non zero
value for $\theta$.  Let us consider the simple case where
\begin{equation}
\theta=\arctan \, (x) ~~ {\rm with} ~~ x \equiv \Omega_1/\Omega_\sigma ~,
\end{equation}
changes from zero to a non-zero value. More concretely, we increase and decrease
$\Omega_1$ by linear ramps such that
\begin{equation}
x(t)= \left\{ \begin{array}{cl} \alpha t \, , & {\rm for} ~ t \in \big(0,
{\textstyle {1 \over 2}} T \big) \\[0.2cm]
\alpha (T-t) \, , & {\rm for} ~ t \in \big(  {\textstyle {1 \over 2}} T,T \big)
\end{array} \right.
\end{equation}
or by a sine ramp given by
\begin{equation}
x(t) = x_{\text{max}} \, \sin \beta t
\end{equation}
with $t \in(0,\pi/ \beta)$. For those two cases one can easily derive the
geometrical phases as functions of the experimental control parameters $T$,
$\alpha$, $x_{\text{max}}$ and $\beta$ and we present the results in Figure
\ref{linear}.

\noindent \begin{minipage}{3.38truein}
\begin{center}
\begin{figure}[ht]
\centerline{
\epsfig{file=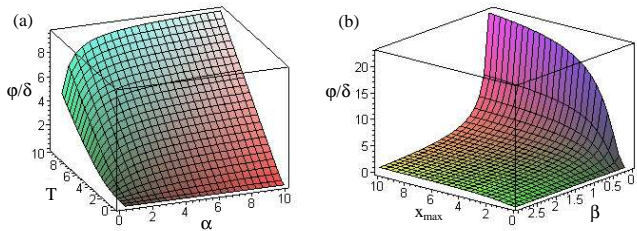,width=3.68truein}
}
\caption[linear]{\label{linear} The resulting geometrical phase
$\varphi^{\text{g}}/\delta$ produced (a) by a linear ramp and (b) by a sine ramp.
The produced geometrical phases are largely independent of the control parameters
$\alpha$ or $x_{\text{max}}$ respectively. $\alpha$ and $\beta$ are
depicted in inverse time units, while $x_{\text{max}}$ is dimensionless.}
\end{figure}
\end{center}
\end{minipage}

From the plots we see that the accumulated geometrical phase $\varphi$ is, as
expected, widely independent on the size of $\alpha$ or $x_{\text{max}}$ giving the
resilience of the proposed E-STIRAP process from control errors, especially from the
exact value of the laser amplitudes or their time derivatives. This independence of
$\varphi$ makes the proposed quantum gate implementation an attractive candidate for
experiments where the exact control of all laser parameters is not possible. In
addition, it is experimentally appealing to construct a setup where the two employed
laser fields can be produced from the same source, separated in two by a beam
splitter and having the ratio $\Omega_1/\Omega_\sigma$ modulated e.g. by an
Acousto-Optical Modulator (AOM). With this setup, the ratio is also resilient to the
fluctuations of the amplitude of the initial laser beam.

\section{Conclusions}\label{conclusions}

In this paper, we discussed in detail concrete proposals for the realization of
universal two-qubit phase gates for quantum computing in atom-cavity setups. The
atoms (or ions) are trapped at fixed positions inside an optical cavity and each
qubit is obtained from two stable ground states of the same atom. To activate a time
evolution, two laser fields are applied simultaneously inducing transitions in an
effectively $\Lambda$-like system. In order to minimize the experimental effort, we
seek quantum computing schemes that are as simple as possible in terms of resources.
The quantum gate implementations proposed here do not require individual addressing
of the atoms and can be realized, essentially, within one step. In particular, we
considered the case where the atom-cavity coupling constant $g$ is only one order of
magnitude larger than each of the spontaneous decay rates of the
system and we assumed $g^2 = 100 \, \kappa \Gamma$.

To avoid dissipation, we populate only the ground state of the system including all
stable ground states of the atoms while the cavity is in its vacuum state.
Nevertheless, an interaction between qubits can take place via virtual population of
the cavity mode and excited atomic levels. We first eliminated the possibility of
leakage of photons through the cavity mirrors. This was achieved by choosing the
parameters such that adiabaticity restricts the time evolution of the system
effectively onto a decoherence-free subspace with respect to cavity decay that
includes highly entangled atomic states. Using these states for the implementation
of Raman and STIRAP-like evolutions, entanglement can be created between the ground
states of the atoms.

Comparing the E-Raman and E-STIRAP procedures we see that they give in general
similar fidelities and success rates, with E-STIRAP having a slight advantage. Their
control procedures are quite different in both cases, providing a flexibility to
adapt the proposed scheme to different experimental setups. The main conceptual as
well as practical difference appears when we produce the entangling phase gates. As
it is seen in Figure \ref{Raman} the phase produced by the E-Raman transition on the
state $|11\rangle$ is given explicitly by the evolution of this state from the north
to the south pole of the Bloch sphere. In contrast, with the E-STIRAP procedure
initial population in the $|11\rangle$ state circulates along a closed path and
returns exactly to the initial point. Nevertheless, at the end the state
$|11\rangle$ acquires a geometrical phase.

The idea of using measurements on ancilla states, like the measurements on the
cavity mode considered in this paper, leads to a realm of new possibilities for
quantum computing. A recent example is the linear optics scheme by Knill {\em et
al.} \cite{Knill} where photons become entangled without ever having to interact. As
long as the measurements on the ancilla state do not reveal any information about
the qubits, they induce a unitary time evolution between them which can then be used
to implement universal gate operations \cite{Sipe}. In contrast to other schemes,
the gate success rate for atom-cavity quantum computing can, at least in principle,
be arbitrarily close to one since the measurements are performed almost
continuously. This allows to utilize the quantum Zeno effect
\cite{beige} which assures that transitions in unwanted states are
strongly inhibited.

The proposed schemes exhibit stability against fluctuations of most system
parameters. The presented dynamical and geometrical phases are obtained by a time
evolution that does not dependent on the exact value of the external control
parameters. Using E-Raman or E-STIRAP processes, we enjoy the experimental
advantages, such as, independence of the final state on the exact intermediate value
of the laser field amplitude. Gates are constructed dynamically as well as
geometrically in order to take advantage of the additional fault-tolerant features
of geometrical quantum computation \cite{Ellinas}. Recently, many attempts have been
made in exploring the decoherence characteristics of geometrical phases
\cite{Bristol,Vedral}. Whether they are advantageous over
equivalent dynamical evolutions depends on the employed physical system. In contrast
to this, we presented here evolutions that are intrinsically decoherence-free and
naturally allow the generation of entangling dynamical as well as geometrical phases
with simple control pictures.

\vspace{0.5cm}
{\em Acknowledgements.} This work was supported in part by the European Union and
the EPSRC. A.B. thanks the Royal Society for a James Ellis University Research
Fellowship.

\end{multicols}
\end{document}